# HARDWARE-ORIENTED GROUP SOLUTIONS FOR HARD PROBLEMS

**Mark Burgin**

Department of Mathematics
University of California, Los Angeles
405 Hilgard Ave.
Los Angeles, CA 90095

## Abstract

Group and individual solutions are considered for hard problems such as satisfiability problem. Time-space trade-off in a structured active memory provides means to achieve lower time complexity for solutions of these problems.

## 1. Introduction

Time of computation is the most popular computational complexity measure and an important parameter, which implies restrictions on practical computability (tractability) in comparison with theoretical/potential computability. In some cases, it is possible to reduce time of a solution of a given problem by means utilization of more memory (space) for computation and applying group solutions, i.e., when the problem is solved for a sequence of parameters. An important peculiarity of this approach is that a structured memory (cf., for example, (Burgin, 2003)) is used to achieve such advantages in time-space trade-off.

Here we do not analyze problems of simulation of computations with a structured memory by means of a traditional Turing machine with one linear tape. These problems will be considered in other works.

In the last section of the paper, individual solutions are considered.

## 2. Boolean expressions

Boolean expressions are built from the following elements:

1. Boolean variables, for example, $x, y, z, x_1, y_3, x_i$ etc.

2. Unary Boolean operator $\neg$ applied to one expression.

3. Binary Boolean operators (symbols of Boolean operations) $\vee$ and $\wedge$ applied to two expressions.

4. Parentheses to group operators and variables, if necessary to alter the default precedence of operators: $\neg$ the highest, then $\wedge$, and finally $\vee$.

Boolean operator (symbols of the Boolean operation) $\neg$ is called *negation*, Boolean operator (symbols of the Boolean operation) $\wedge$ is called AND or *conjunction*, and Boolean operator (symbols of the Boolean operation) $\vee$ is called OR or *disjunction*.

Boolean expressions are built by an inductive process.

**Definition 1**. 1. Any Boolean variable is a Boolean expression.

2. If *A* is a Boolean expression, then ¬*A* and ¬(*A*) are also Boolean expressions.

3. If *A* and *B* are Boolean expressions, then *A*∨*B*, *A*∧*B*, (*A*) ∨ (*B*), and (*A*) ∧ (*B*) are also Boolean expressions.

Binary Boolean operators ∨ and ∧ are associative. Thus, they define integral operations in the sense of (Burgin and Karasik, 1976). We denote these integral operations, which are applied to an arbitrary number of Boolean variables, by the same symbols: ∨ and ∧.

Boolean variables take two values 1 and 0, which are traditionally called truth values: 1 denoting *True* and 0 denoting *False*. *Truth values* of Boolean expressions, which are equal either to 1 or to 0, are defined by truth tables for Boolean operators and by an inductive process when truth values of all variables in this expression are given.

**Definition 2.** The set *T* of truth values for all variables in a Boolean expression *A* is called a *truth assignment* for this expression.

When *T* is given, we define the value *A*(*T*) of the Boolean expression *A*. For instance, taking the Boolean expression (*x*∧¬*x*∧*z*∧¬*y*∧*y*)∨(*u*∧¬*x*∧*z*∧¬*y*∧*w*) and the truth assignment *T* = { *u* = 1, *x* = 0, *z* = 1, *y* = 0, *w* = 1}, we have *A*(*T*) = 1.

**Definition 3.** A truth assignment *T* for a Boolean expression *A* *satisfies A* if *A*(*T*) = 1, i.e., the makes expression *A* true.

**Definition 4.** A Boolean expression *A* is called *satisfiable* if there is a truth assignment *T* for *A* that satisfies *A*. For instance, *x*∨¬*x*, is satisfiable, while *x*∧¬*x*, is not satisfiable.

There are two normal forms of Boolean expressions: disjunctive and conjunctive. They are defined through the following concepts.

**Definition 5.** A *literal* is either a Boolean variable, e.g., *x*, or a negated Boolean variable, e.g., ¬*x*.

**Definition 6.** A *disjunctive clause* is a literal or the conjunction of two or more literals, e.g., *x*∨¬*x*∨*z*∨¬*y*∨¬*u*.

**Definition 7.** A *conjunctive clause* is a literal or the disjunction of two or more literals, e.g., $x \wedge \neg x \wedge z \wedge \neg y \wedge y$.

**Definition 8.** A Boolean expression $A$ is in the *disjunctive normal form* or DNF if it is the disjunction of conjunctive clauses, e.g., $(x \wedge \neg x \wedge z \wedge \neg y \wedge y) \vee (u \wedge \neg x \wedge \neg z \wedge \neg y \wedge w) \vee (x \wedge u \wedge w \wedge v \wedge y)$.

**Definition 9.** A Boolean expression $A$ is in the *conjunctive normal form* or CNF if it is the conjunction of disjunctive clauses, e.g., $(x \wedge \neg x \wedge z \wedge \neg y \wedge y) \vee (u \wedge \neg x \wedge \neg z \wedge \neg y \wedge w) \vee (x \wedge u \wedge w \wedge v \wedge y)$.

**Lemma 1**. For any Boolean expression $A$ there is an equivalent Boolean expression $D$ in the disjunctive normal form.

**Lemma 2**. For any Boolean expression $A$ there is an equivalent Boolean expression $C$ in the conjunctive normal form.

It is possible to find proofs, for example, in (Davis and Weyuker, 1983).

## 3. Group Solutions

An algorithmic problem $P$ is usually a question about some properties of objects from an infinite, as a rule, set $X$. The most popular in theoretical literature algorithmic problem is the question if a given word $u$ belongs to a given language $L$, which, in this case, is the set $X$.

**Definition 10.** Solving an algorithmic problem $P$ for one object will be called an *individual solution*.

**Definition 11.** Solving an algorithmic problem $P$ for a group of objects from $X$ will be called a *group solution*.

We assume that the set $X$ is countable and an evaluation function $l: X \to N$ into the set $N$ of all whole numbers is defined. For instance, any enumeration of $X$ is such an evaluation function.

**Definition 12.** Solving an algorithmic problem $P$ for all objects $u$ from $X$ with $l(u) \leq n$ for some number $n$ will be called a *sequential solution*.

We will show how a sequential solution allows one to decrease time for each individual solution. To do this, we take as an evaluation function on the set **B** of all Boolean expressions the length $l(A)$ of a Boolean expression $A$ and consider the popular *satisfiability problem* SAT:

Given a Boolean expression $A$, find if it is satisfiable.

**Remark 1.** It is possible to consider several definitions of the length $l(A)$ of a Boolean expression $A$: a) $l(A)$ is equal to the number of symbols in $A$, that is number of Boolean variables, Boolean operations, and parentheses; b) $l(A)$ is equal to the number of Boolean variables in $A$; c) $l(A)$ is equal to the number of distinct Boolean variables in $A$; d) $l(A)$ is equal to the number of symbols in the coding of $A$, as usually Boolean variables are coded (for example, $x_5$ is coded by $x11111$, cf., for example, (Davis and Weyuker, 1983)) in order to have a finite alphabet $\Sigma$ for their representation.

**Theorem 1.** There is a sequential solution for SAT such that each individual solution is obtained in polynomial time.

Proof. As we are interested in practical aspects of computation, we do not use Turing machines for building such algorithm, although it is possible to simulate all described procedures and operations by a conventional Turing machine. SAT is considered here for all Boolean expressions represented in a finite alphabet $\Sigma$ with $m$ symbols. The computational structure utilized here consists of a (potentially infinite) number of copies of the switching element $\mathbf{SW}_m$, memory cells that store symbols from the alphabet $\Sigma$, a separating structure $\mathbf{D}$, and three finite automata $\mathbf{A}_{neg}$, $\mathbf{A}_{dis}$, and $\mathbf{A}_{con}$. It is possible to consider such structure as a grid automaton (Burgin, 2003a) or realize it as a kind of structured memory, an active structured memory containing elements that establish connections between cells in the process of inductive Turing machine functioning (cf., for example, (Burgin, 2003)).

The switching element $\mathbf{SW}_m$ has one inlet and $m$ outlets each of which corresponds to one symbol from $\Sigma$. It works in the following manner. A symbol $a$ from $\Sigma$ is given to $\mathbf{SW}_m$ as an input. After this, $\mathbf{SW}_m$ activates (opens) its outlet that corresponds to the symbol $a$. It is possible to assume that it takes one unit of time to perform a switching or if we build $SW_m$ using Boolean elements, time of switching is equal to $m$. Such switching elements are described in (Minsky, 1967).

The automata $A_{neg}$, $A_{dis}$, and $A_{con}$ realize Boolean operations negation, disjunction, and conjunction, correspondingly. Such simple automata are called Boolean elements (Minsky, 1967).

The separating structure $D$ can separate a list of words (items) into two parts, given a condition for separation. For instance, by Definition 1, each Boolean expression $H$ with $l(A) > 1$ has one of the following forms: $\neg A$, $\neg(A)$, $A \vee B$, $A \wedge B$, $(A) \vee (B)$, or $(A) \wedge (B)$. In the first two cases, the separating structure $D$ can separate $A$ and $\neg$. In all other cases, the separating structure $D$ can separate $A$ and $B$. It is possible to realize such separation structure $D$ by a simple Turing machine that demands $O(n)$ operations to do separation. To optimize this procedure and consequent operations of comparison, it is better to write one list on one tape and another list on the second tape. Utilization of Turing machines with many tapes does not change polynomial time of computations (Hopcroft, Motwani, and Ullman, 2001).

We correspond to each Boolean expression $A$ with the length $l(A) < n$ a cell $c_A$ of the memory. When it found whether $A$ is satisfiable or not, symbol 1 is stored in the cell $c_A$ in the first case, symbol 0 in the second case, and symbol $t$ when $A$ is a *tautology*, i.e., $A$ is satisfiable by all truth assignments.

The concept of tautology brings us to another algorithmic problem called *tautology problem* TAU:

Given a Boolean expression $A$, find if it is a tautology.

In addition, a network $N$ of switching elements $SW_m$ is built so that it is possible to come to the cell $c_A$ making $l(A)$ steps when $A$ is given as an input to this and the switching element $SW_m$ performs switching in one unit of time (in one step). When, as it was explained above, the switching element $SW_m$ needs $m$ units of time to perform switching, then reaching the cell $c_A$ demands $m \cdot l(A)$ steps where $m$ is a fixed number.

We prove Theorem 1 by induction in the length of Boolean expressions, finding sequential solution to two problems at the same time: SAT and TAU.

Let $l(A) = 1$. Then $A$ contains only one Boolean variable and we can check if it is satisfiable and if it is a tautology in two steps.

Let $l(A) = n + 1$ and assume that for all Boolean expressions $D$ with $l(D) < n + 1$ both problems, SAT and TAU, are solved, the cells $c_D$ are correctly filled, and the network **N** leading to these cells is built. By Definition 1, $A$ is equal either to $\neg D$ or to $\neg(D)$ or to $D \vee H$ or to $D \wedge H$ or to $(D) \vee (H)$ or to $(D) \wedge (H)$. Then in a polynomial numbers of steps (polynomial time), we can reach the cells $c_D$ and $c_H$, finding whether $D$ and $H$ (in the first two cases, only $D$) are satisfiable, tautologies or not satisfiable.

To find satisfiability of $A$, we use the following properties of Boolean expressions.

$\neg D$ is satisfiable if and only if $D$ is not a tautology.

$D \vee H$ is satisfiable if and only if either $D$ or $H$ or both are satisfiable.

These properties allow to find satisfiability of $A$ using the automata $A_{neg}$, $A_{dis}$. It demands a fixed number of steps, which does not depend on the length of $A$.

In the case when $A$ is equal to $D \wedge H$, it follows from De Morgan's laws that $A$ is equivalent to the expression $\neg(\neg D \vee \neg H)$. As $l(D) < n + 1$ and $l(H) < n + 1$, we can check satisfiability of $A$ in the same way as in two previous cases.

In a similar way, we check if $A$ is a tautology for $A$ is equal either to $\neg D$ or to or to $D \wedge H$, utilizing the following properties of Boolean expressions.

$D \wedge H$ is a tautology if and only if both $D$ and $H$ are tautologies.

As $D$ is equivalent to $\neg\neg D$, we also have:

$\neg D$ is a tautology if and only if $D$ is not satisfiable.

For the case when $A$ is equal to $D \vee H$, we use the equivalent formula $\neg(\neg D \wedge \neg H)$.

The principle of induction concludes the proof of Theorem 1.

This proof also gives us the following result.

**Theorem 2.** There is a sequential solution for TAU such that each individual solution is obtained in polynomial time.

**Remark 2.** The network **N** built for sequential solutions in the proof of Theorem 1 grows very fast. Actually, it has exponential speed of growth. As a result Theorem 1 demonstrates how the time-space trade-off allows one to achieve very high speed of computation utilizing very big memory. This technique is similar to the technique used in (Burgin, 1999) to demonstrate that inductive Turing machines with a structured memory are more efficient than conventional Turing machines.

Theorem 1 allows us to reconsider the problem of relations between classes **P** and **NP**.

**Theorem 3** (Cook Theorem, 1971)**.** SAT is **NP**-complete.

Now it is possible to find a proof of this theorem in many textbooks (cf., for example, (Davis and Weyuker, 1983) or (Hopcroft, Motwani, and Ullman, 2001)).

Cook's Theorem means that all problems in NP can be reduced to SAT.

**Theorem 4** (Cook-Levin Theorem)**.** SAT is in **P** if and only if **NP** = **P**.

Now it is possible to find a proof of this theorem in some textbooks (cf., for example, (Sipser, 1997)).

Theorems 1 and 4 imply the following result.

**Theorem 5.** With respect to sequential solutions, **NP** = **P**.

### 4. Individual Solutions

At first, we consider Boolean expressions in the disjunctive normal form.

**Lemma 3**. A conjunctive clause is satisfiable if and only if it does not contain a Boolean variable and its negation.

Indeed, expression $x \wedge \neg x$ is not satisfiable. Consequently, any conjunctive clause that contains this expression is not satisfiable. In the case when a conjunctive clause $C$ contains only different variables, we take the following truth assignment $T$: 1 is assigned to all variables without negation, while 0 is assigned to all variables with negation. By the definition of conjunction, this truth assignment satisfies $C$.

Dual to Lemma 3 is the following result.

**Corollary 1**. A conjunctive clause is not satisfiable if and only if it contains some Boolean variable and its negation.

The definition of disjunction implies the following result.

**Lemma 4**. Boolean expression $A$ is in the disjunctive normal form is satisfiable if and only if at least one of its conjunctive clauses is satisfiable.

These results allow us to treat a restricted version of the satisfiability problem SAT that is called the *disjunctive satisfiability problem* DSAT:

Given a Boolean expression $A$ in the disjunctive normal form, find if it is satisfiable.

As above, DSAT is considered here for all Boolean expressions represented by words in a finite alphabet $\Sigma$ with $m$ symbols.

**Theorem 6.** There is an individual solution for DSAT obtained in polynomial time.

Proof. Let us consider a Boolean expression $A$ in the disjunctive normal form that contains $m$ conjunctive clauses $C_1$, $C_2$, ... , $C_m$ and $n$ Boolean variables. In each of these clauses, all variables without negation go at the beginning before all variables with negation and variables in both groups are ordered according to their indices. We show that there is a polynomial $p(n)$ such that it takes not more than $p(r)$ steps/operations to find if an arbitrary conjunctive clause $C$ is satisfiable where $r$ is the number of Boolean variables in $C$. By Lemma 3, it is sufficient to find if $C$ does not contain a Boolean variable and its negation.

To continue, we prove the following statement (E).

Finding two equal variables in two lists $L = \{ a_1, a_2, ..., a_k \}$ and $M = \{ b_1, b_2, ..., b_h \}$ of variables $x_i$ or asserting that $L$ and $M$ have no equal variables demands $r$ operations of comparison with $r \leq 2(d(L) + d(M))$ where there are no equal elements in each list and $d(L)$ and $d(M)$) are numbers of items in $L$ and $M$, correspondingly. We assume that all variables $x_i$ are ordered by their indices, that is, $x_i < x_j$ if $i < j$. This is not a limitation because there are sorting algorithms (for example, the QUICK SORT, HEAP SORT, and MERGE SORT) that have time complexity of $O(n \log n)$ (cf., for example, (Knuth, 1973)).

We prove it by induction in $n$ that is equal to the sum of items in $L$ and $M$, i.e., $n = d(L) + d(M)$.

The base of induction.
1. When $d(L) = 0$ or $d(M) = 0$, then $r = 0$ and the statement (E) is true.
2. When $d(L) = d(M) = 1$, then $r = 1$ and the statement (E) is true.

3. When $d(L) = 1$ and $d(M) = 2$, then it is sufficient to compare $a_1$ with $b_1$ and $b_2$ ; thus, $r = 2 \leq 1 + 2 = 3$ and the statement (E) is true.

4. When $d(L) = 2$ and $d(M) = 1$, everything is symmetric to the previous case.

A step of induction.

Assume that the statement (E) is true for $n - 1$, there are two lists $L = \{ a_1, a_2, \ldots, a_k \}$ and $M = \{ b_1, b_2, \ldots, b_h \}$, $1 < h \leq k \leq n - 1$ and a new element $a$ is added to $L$, giving the new list $L_1$. As there are no equal elements in each list, there are three possible cases: (1) $a < a_1$; (2) $a_{l-1} < a < a_l$; (3) $a_k < a$.

Case 1. Comparison of elements from two lists starts with $a$. It is compared consecutively with $b_1, b_2, \ldots$ If it found that $a = b_i$ for some $i \leq h$, then $r = i \leq h < 2(k + 1 + h)$ and the statement (E) is true for $n$.

If it found that $a > b_h$, then $a_i > b_h$ for all $i$, $= 1, 2, \ldots, k$ and it takes $r = h$ comparisons to find this. Thus, $r = h < 2(k + 1 + h)$ and the statement (E) is true.

If it found that $a < b_i$ for some $1 \leq i \leq h$, then it takes $i$ comparisons to find this. After this, we have two lists $L$ and $M_1$ to compare and find equal elements where $M_1 = \{ b_i, b_{i+1}, \ldots, b_h \}$. For them, we have $d(L) = k$ and $d(M_1) = h - i + 1$. By the induction assumption, it takes $r_1$ operations of comparison to find two equal variables in these lists where $r_1 \leq 2(d(L) + d(M_1)) = 2(k + h - i + 1)$ and $1 \leq i \leq h$. Thus, it takes $r$ operations of comparison to find two equal variables in lists $L$ and $M$ where $r = r_1 + i$. Consequently, we have $r \leq 2(k + h - i + 1) + i = 2k + 2h - 2i + 2 + i = 2k + 2h + 2 - i = 2(k + h) + 2 - i = 2(d(L_1) + d(M)) + 2 - i$ and the statement (E) is true when $i \geq 2$.

Now what does it mean that $i = 1$? It means that $a < b_1$ and we need only one comparison to find this. After this, we have two lists $L$ and $M$ to compare and find equal elements. By the induction assumption, this demands $q$ operations of comparison where $q \leq 2(k + h)$. Then $r = q + 1 \leq 2(k + h) + 1 \leq 2(k + h + 1) = 2(d(L_1) + d(M))$ and the statement (E) is true for $n$.

This completes the proof for the case 1.

It is possible to reduce cases 2 and 3 to the case 1, taking $a_1$ as $a$. Thus, by the principle of induction, the statement (E) is true for all $n$.

It is necessary to remark that when variables $x_i$ are coded as words in the alphabet $\Sigma$, then each comparison demands $O(\log_2 n)$ steps. However, in this case the length of Boolean expressions is also multiples by the same factor. As a result, coding does not violate polynomial time of computation.

Statement (E) implies that there is a polynomial $p(n)$ such that it takes not more than $p(r)$ steps/operations (where $r$ is the length $l(C)$ of $C$, i.e., the number of Boolean variables in $C$) to find if an arbitrary conjunctive clause $C$ is satisfiable because there is linear in time algorithm that separates all variables without negation and all variables with negation in $C$, making two lists for comparison.

If there is a polynomial $p(n)$ such that it takes not more than $p(r)$ steps/operations to find if an arbitrary conjunctive clause $C$ is satisfiable, then for some fixed number $d$ it is possible to check satisfiability of an arbitrary conjunctive clause $C$ in time (number of steps) that is less than $r^d$.

Now having a Boolean expression $A$ in the disjunctive normal form that contains $m$ conjunctive clauses $C_1, C_2, \ldots, C_m$ and $n$ Boolean variables, we can check satisfiability of $A$ by successively checking satisfiability of the conjunctive clauses $C_1, C_2, \ldots, C_m$. If the procedure for the clause $C_i$ demands $q_i$ steps, then the procedure for the whole $A$ demands $q = q_1 + q_2 + \ldots + q_m + k$ steps where $k$ is a fixed number. Then $q < l(C_1)^d + l(C_2)^d + \ldots + l(C_m)^d + k < (l(C_1) + l(C_2) + \ldots + l(C_m))^d$ when we take sufficiently big $d$.

This concludes the proof of Theorem 1.

Another restricted version of the satisfiability problem SAT is called the *conjunctive satisfiability problem* CSAT:

Given a Boolean expression $A$ in the conjunctive normal form, find if it is satisfiable.

De Morgan's laws, Theorem 6, and a possibility to go from DNF to CNF in a polynomial time imply the following result.

**Theorem 7.** There is an individual solution for DSAT obtained in polynomial time.

As it is proved that it is possible to reduce SAT to CSAT in a polynomial time, we have the following result.

**Theorem 8** (Cook's Theorem, 1971). CSAT is **NP**-complete.

Now it is possible to find a proof of this theorem in many textbooks (cf., for example, (Hopcroft, Motwani, and Ullman, 2001)).

Theorems 6 and 4 imply the following result.

**Theorem 9. NP** = **P**.

**5. Conclusion**

Here very rough estimates of computational time are given. Later research will give more exact time and space characteristics.